# The Impact of Mechanical Strain on Magnetic and Structural Properties of 2D Materials: A Monte Carlo study

Aytac Celik
Metallurgical and Materials Engineering Department,
Sinop University, Sinop, 57000
Turkey
Corresponding author: Aytac Celik, aytaccelik@gmail.com

**Abstract**

The inherent flexibility of two-dimensional (2D) materials allows for efficient manipulation of their physical properties through strain application, which is essential for the development of advanced nanoscale devices. This study aimed to understand the impact of mechanical strain on the magnetic properties of two-dimensional (2D) materials using Monte Carlo simulations. The effects of several strain states on the magnetic properties were investigated using the Lennard-Jones potential and bond length-dependent exchange interactions. The key parameters analyzed include the Lindemann coefficient, radial distribution function, and magnetization in relation to temperature and magnetic field. The results indicate that applying biaxial tensile strain generally reduces the critical temperature ($T_c$). In contrast, the biaxial compressive strain increased $T_c$ within the elastic range, but decreased at higher strain levels. Both compressive and tensile strains significantly influence the ferromagnetic properties and structural ordering, as evidenced by magnetization hysteresis. Notably, pure shear strain did not induce disorder, leaving the magnetization unaffected. In addition, our findings suggest the potential of domain-formation mechanisms. This study provides comprehensive insights into the influence of mechanical strain on the magnetic behavior and structural integrity of 2D materials, offering valuable guidance for future research and advanced material design applications.

**Word count:** 5,083 words, excluding references.
**Funding Statement:** This research received no external funding.
**Ethical Compliance:** This study was conducted in accordance with all relevant ethical guidelines and standards. Because the study did not involve human participants or animals, ethical approval was not required. The authors declare that there are no conflicts of interest, and that all data and materials comply with relevant regulations and ethical standards.
**Data Access Statement:** Research data supporting this publication are available upon request.
**Conflict of Interest Declaration:** The authors have no conflicts of interest to disclose.
**Author Contributions:** The author was solely responsible for all aspects of this work, including the conceptualization, methodology, investigation, data analysis, and writing of the manuscript.

**Plain Language Summary**

This study investigates the effects of different types of mechanical strain on the magnetic and structural properties of thin two-dimensional materials using Monte Carlo simulations. Three types of strains were examined: stretching in all directions (biaxial tensile strain), compression in all directions (biaxial compressive strain), and twisting (pure shear strain). Our findings showed that stretching generally lowers the temperature at which a material becomes magnetically ordered. However, compression initially raises this temperature, but lowers it again if the compression is too high. The twisting did not affect the magnetic properties of the material. These insights help us to understand how to design better materials for advanced technologies.

**Introduction**

The effects of mechanical strain on the magnetic and structural characteristics of 2D materials are of interest in Physical Science. The ultrathin and flexible structure of 2D materials enables efficient manipulation of physical properties by applying strain. Strain-coupled multiferroic heterostructures are crucial for the development of energy-efficient nanoscale devices that can be controlled using voltage, as highlighted by Ji et al. (2023)[1]. However, these materials experience spontaneous and unpredictable deformations, resulting in a localized strain, which makes their consistent use difficult.

Recent progress has focused on addressing these limitations through the development of methods for controlled deformation. This includes utilizing mismatched thermal expansion coefficients and pre-stretched or pre-patterned substrates to achieve specific strain distributions [2–4]. Understanding the relationship between the mechanical strain and magnetic behavior is essential for the advancement of sophisticated technological applications that require customized features in 2D magnetic materials.

Theoretical models suggest that 2D materials such as graphene have the potential to withstand significant elastic strains of up to 30%. However, in practical applications, the strain levels achieved are typically significantly lower (< 5%) [5]. This disparity underscores the difficulties and possibilities for future research to close this divide and to expand the range of applications of these materials in electronics. In addition, scientific and applied interest in magnetoelastic (ME) coupling behavior in these heterostructures has increased significantly.

Significant progress has been made in the magnetic characterization of 2D materials through numerous modeling and simulation studies. In their study, Ye, Sun, and Jiang (2015) created a vibronic Ising-like model that includes harmonic stretching and bending interactions. This model was used to simulate the properties of the spin-crossover (SCO) materials more realistically [6]. Frechette, Dellago, and Geissler (2020) provided additional clarification on the source of mean-field behavior in elastic Ising models. They accomplished this by formulating an effective Hamiltonian that accurately represented the thermodynamics and kinetics of spin-crossover systems. Furthermore, they extended their investigation to include nanocrystals [7]. In addition to these findings, a further investigation conducted by Frechette et al. (2021) showed how elastic forces contribute to the creation of non-equilibrium patterns in the process of nanocrystal ion exchange [8]. Schebarchov, Schulze, and Hendy (2014) introduced a degenerate Ising model to simulate crystal-melt interfaces, which accurately represents the coexistence of different phases and the impact of interface topology on phase transitions[9]. Belim et al. (2022) examined the effect of different substrates on the Curie temperature of 2D magnets. They used a model that combines the Ising and Frenkel-Kontorova potentials to demonstrate how deformations caused by the substrate can greatly influence the magnetic properties of these magnets [10,11].

However, a significant knowledge gap exists in understanding how different types of mechanical strain (biaxial tensile, compressive, and pure shear) specifically affect the magnetic properties and structural ordering of 2D materials. Previous studies have not thoroughly explored the threshold levels of strain that lead to changes in domain formation and critical temperature, leaving important aspects of strain-induced phenomena unexplored.

Moreover, strain engineering using techniques such as thin-film stressor deposition offers precise control over the magnitude and directionality of the strain, and whether it induces compression or tension[12]. This control is crucial for investigating low-temperature phenomena including strain-influenced phase transitions, which can lead to innovative applications in nanoelectronics and quantum computing [13]. Despite the progress in understanding the effects of mechanical strain on the magnetic properties of 2D materials, there remains a need for a fundamental understanding of the interplay between strain and magnetic behavior in various scenarios.

Theoretical studies on the compressible Ising model, as introduced by Baker and Essam (1970, 1971)[14,15], have laid the foundation for understanding the effects of lattice compressibility on critical behavior. Their work focused on a small-displacement approximation, where the coupling constant linearly depended on the interatomic distances, using harmonic bonds to describe the interactions analytically. This approach provides significant insights into the phase transitions and critical behavior of compressible systems, particularly in materials such as β-brass. Gunther, Bergman, and Imry (1971, 1973)[16,17] further extended this model by exploring the effects of applied strain and clarifying the renormalized critical behavior of the system. They provided a more nuanced understanding of the compressible Ising model, highlighting the delicate balance between first- and second-order phase transitions under varying conditions. Small-displacement approximation with the harmonic model limits its ability to accurately capture nonlinear atomic interactions at high deformation rates or strains. In contrast, modern extensions of this model account for non-symmetric forces by considering potentials such as the Lennard-Jones potential, which exhibits both repulsive and attractive components. This allows for more realistic simulations of material behavior, including how they respond to thermal vibrations or high tensile/compressive stresses.

This study seeks to address this knowledge gap by investigating the effects of mechanical strain on the magnetic and structural characteristics of 2D materials using controlled deformation techniques, as detailed in the Methodology section. The investigation was conducted using Monte Carlo simulation of the dynamic Ising model. This paper presents computer simulations of a dynamic Ising model applied to strained 2D magnetic materials. The focus was on examining the changes in critical temperature, coercivity, and domain formation. To simplify the process, a simple model containing an isotropic Lennard-Jones potential and bond length-dependent exchange interactions was built and simulated using approximately 1000 atoms. Although this method compromises some precision compared with systems with an order of magnitude higher number of atoms, it dramatically reduces computational costs, allowing a large number of tests to be run.

**Methods**

To examine the impact of strain on magnetic behavior, multiple crucial components were analyzed. The lattice structure functions as the primary framework, establishing the initial locations of atoms and enabling the measurement of deformation and temperature effects[18]. The elastic energy is crucial for quantifying the material deformation under an applied strain, revealing the impact of mechanical forces on the physical and magnetic properties. The Lennard-Jones potential is essential for modeling isotropic interactions between atoms, capturing both attractive and repulsive forces, thus affecting the mechanical stability and response to strain and thermal vibrations [19,20]. The magnetic Hamiltonian, which incorporates exchange interactions that vary with distance, provides an accurate representation of the interactions between magnetic moments in neighboring atoms. This is essential for precisely anticipating alterations in magnetic characteristics caused by strain-induced effects [21].

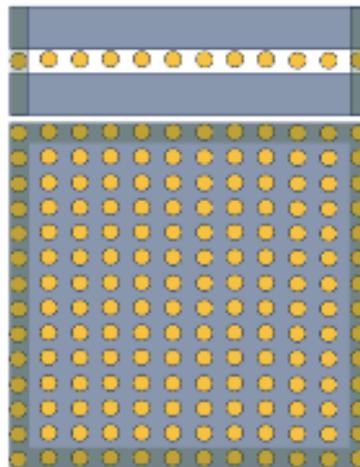

*Figure 1* Top and side views of the simulated system with N atoms in the plane, where boundary atoms with fixed positions are indicated by the dark border.

Initially, the N atoms were arranged in $L_x \times L_y$ box with a square-lattice structure. The lattice structure was constructed using cubic Bravais lattice vectors $R = R_1 a_1 + R_2 a_2$, where $a_1 = a_A x_1$ and $a_2 = a_A x_2$ are the basis vectors with the lattice constants used in this study. This configuration defines the lattice positions at rest in the initial setup (Figure 1) [22]. In the simulations, the atomic lattice (Figure 1) was uniformly stretched, after which the system was relaxed under the influence of the Lennard–Jones potential to determine the equilibrium atomic positions. This relaxation was performed using the Monte Carlo method with fixed boundary conditions, where the positions of the atoms at the edges of the simulation box were held constant. The relaxed atomic positions were then used to calculate the exchange interactions within the Ising model, which were simulated using free-boundary conditions.

The Lennard-Jones potential was employed to model the isotropic interactions among atoms [23], crucial for comprehending both the attractive and repulsive forces in 2D materials.

$$V(r_{ij}) = 4\epsilon \left[ \left(\frac{\sigma}{r_{ij}}\right)^{12} - \left(\frac{\sigma}{r_{ij}}\right)^{6} \right] \qquad (1)$$

where, $\epsilon(=10)$ represents the depth of the potential well, $r_{ij}$ refers to the interatomic distances and $\sigma(=1)$ is the finite distance at which the inter-particle potential is zero. The Lennard-Jones potential, a model that simulates the forces between two particles, captures the key features of interatomic interactions and provides a foundation for analyzing the mechanical properties of a material under different strain and thermal conditions.

The Monte Carlo algorithm was used to study the temperature fluctuations of the atoms around a lattice by maintaining a constant temperature, volume, and number of particles. In this method, atoms are initially positioned on a lattice and random displacements are proposed. The energy change from each displacement determines whether the move is accepted according to the Metropolis criterion, which ensures that the system samples the configurations according to the Boltzmann distribution at the target temperature. The Monte Carlo move method proposes a random small displacement of atoms from their current positions. The change in energy $\delta E_{LJ}$ resulting from the displacement was calculated using the Lennard-Jones potential. If the energy decreases ($\delta E_{LJ} < 0$), then the move is accepted. If the energy increases ($\delta E_{LJ} > 0$), accept the move with probability P = $exp(-\delta E_{LJ}/k_B T)$, where $k_B$ is the Boltzmann constant. This allows the system to occasionally accept higher-energy states, enabling it to effectively sample the full phase space.

Exchange interactions result from combining the Pauli Exclusion Principle with Coulombic interactions between electrons. The Pauli Exclusion Principle asserts that it is impossible for two electrons in an atom to have identical quantum numbers. Consequently, electrons with identical spins should exhibit distinct spatial distributions. This results in an efficient force between electrons, which is not only based on Coulomb's law but is also referred to as the exchange interaction. The energy of the system is influenced by the interactions between the two electrons, particularly their electrostatic repulsion (Coulomb force), which results in the symmetry of the wave function (and hence, the overall state of the system) being dependent on the distance between them[24]. To model the magnetic properties of 2D materials, a magnetic Hamiltonian with distance-dependent exchange interaction parameters was incorporated to accurately describe the coupling between the spin moments of neighboring atoms in the first-degree approximation[10,11].

$$J(r_{ij}) = J_0 \exp\left(-\frac{r_{ij}-r_o}{r_o}\right) \qquad (2)$$

The function $J(r_{ij})$ accounts for the decay in the interaction strength with $J_0(=1)$ exchange interaction parameter, interatomic distances $(r_{ij})$ and ideal bond length $r_o(=1)$, which are crucial for understanding the strain-induced changes in magnetic properties. The Hamiltonian for magnetic properties, assuming nearest-neighbor Ising interactions as a function of the distances between nearest neighbors, is written as:

$$H = - \sum_{ij}^{N} J(r_{ij}) S_i S_j - \sum_{i}^{N} h\, S_i \qquad (3)$$

where $J(r_{ij})$ is the exchange interaction parameter, which depends on the distance $r_{ij}$ between the ith and jth spins; $S_i$ and $S_j$ are the spin variables, which can take values $\pm 1$, and h denotes the external magnetic field applied to the system. In the Hamiltonian, strain is incorporated through the Lennard-Jones potential, which determines the equilibrium interatomic distances. These strain-induced modifications in atomic spacing directly influence the exchange interaction term within the Hamiltonian. This approach ensures that the Hamiltonian accurately captures the energy of a system under the influence of a strain. Therefore, the interplay between strain and magnetic properties in 2D materials can be understood through the modification of exchange interactions and magnetic anisotropy [25]. The Ising model, which incorporates strain-dependent exchange interactions, offers a theoretical framework for predicting these processes and for directing experimental studies.

For Monte Carlo (MC) Ising simulations, the standard Metropolis acceptance rate with the canonical ensemble was used. A standard spin-exchange mechanism was employed for magnetic spin dynamics. In this scheme, an atom is randomly chosen to execute one step of the system evolution and the energy change is calculated. If this energy change is negative, exchange is performed; if it is positive, exchange is performed with the probability $exp(-\delta E/k_B T)$. The Monte Carlo time was measured using one sweep (MCS) after the thermal equilibration of the atoms. Each simulation begins after lattice points are created, and a random spin of atoms, where the spins of atoms are randomly chosen to be -1 or +1 with 50% probability, is defined. Subsequently, the desired strain was applied to the system. The system was equilibrated at a temperature ($T = 3.5$) above the Curie temperature ($T_c$). After equilibration with $4x10^4$ MCS, the configurations were recorded and analyzed at each MC step, new state parameters were defined, and the calculations were repeated[26].

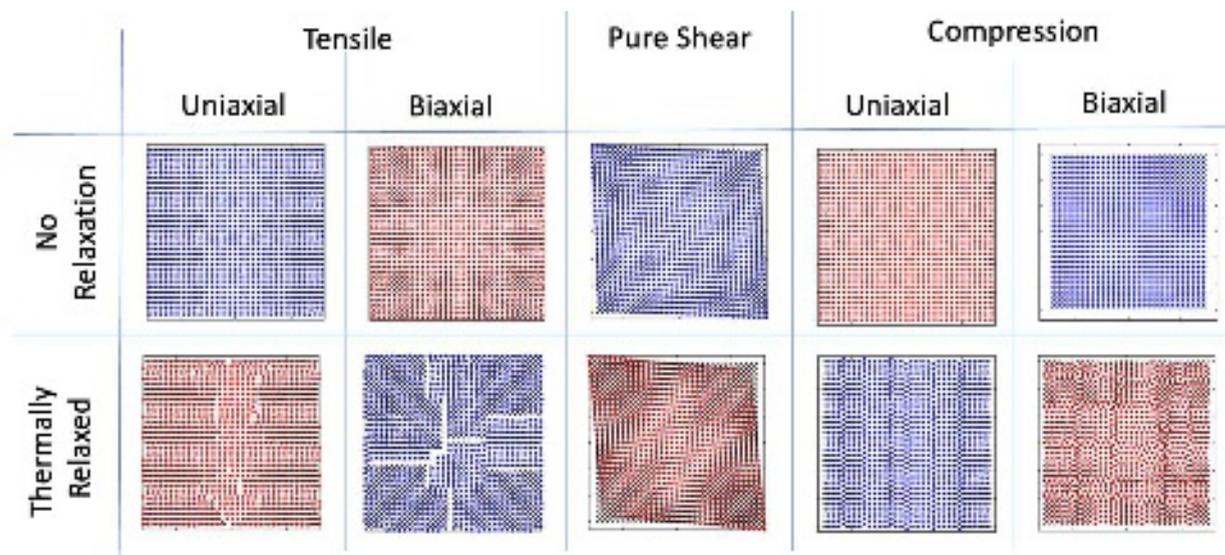

*Figure 2* 2D magnetic materials under the strain conditions

Test simulations showed that a system with 1000 atoms and 4000 MCS was sufficient to demonstrate the effect of strain on the magnetic behavior of 2D materials. For temperature and magnetic field sweeps, 150 steps were used to span the interval. For thermal relaxation, $6\times10^6$ MCS and magnetic analysis of $1.2\times10^7$ MCS were performed for each simulation. Each parameter set experiment was repeated 10 times to obtain the average behavior of the system for 1000 simulations. It must be noted that larger system sizes are getting exponentially longer times and larger file sizes owing to additional thermal and Leonard-Jones calculations. Simulations performed with several strain fields revealed distinctive thermal behaviors with a significant impact on the magnetic and thermodynamic properties (Figure 2). A few physical differences between the simulation results are shown in Figure 2, where the atomic positions can be observed at the lowest simulated temperature (T =0.1). At high tensile strains, cryogenic cracking occurs, which affects the thermodynamic values. At high compressive strains, domains featuring shear walls were created, resembling the nonuniform strain distributions observed in misfit-strained 2D materials. This phenomenon is comparable to the buckling and wrinkling of two-dimensional (2D) materials.

**Results**

This study examines the impact of strain on the magnetic characteristics of 2D materials by specifically analyzing thermally relaxed and non-relaxed systems. Various strain conditions were utilized to investigate their impact on stability, magnetic transitions, and structural order. The analysis comprised measurements of the Lindemann coefficient [27], a histogram of bond lengths, magnetization as functions of temperature and magnetic field, and a radial distribution function. For brevity, the results of all the simulations with uniaxial strains are not presented herein. However, they demonstrated patterns comparable to those observed for the biaxial and pure shear strains.

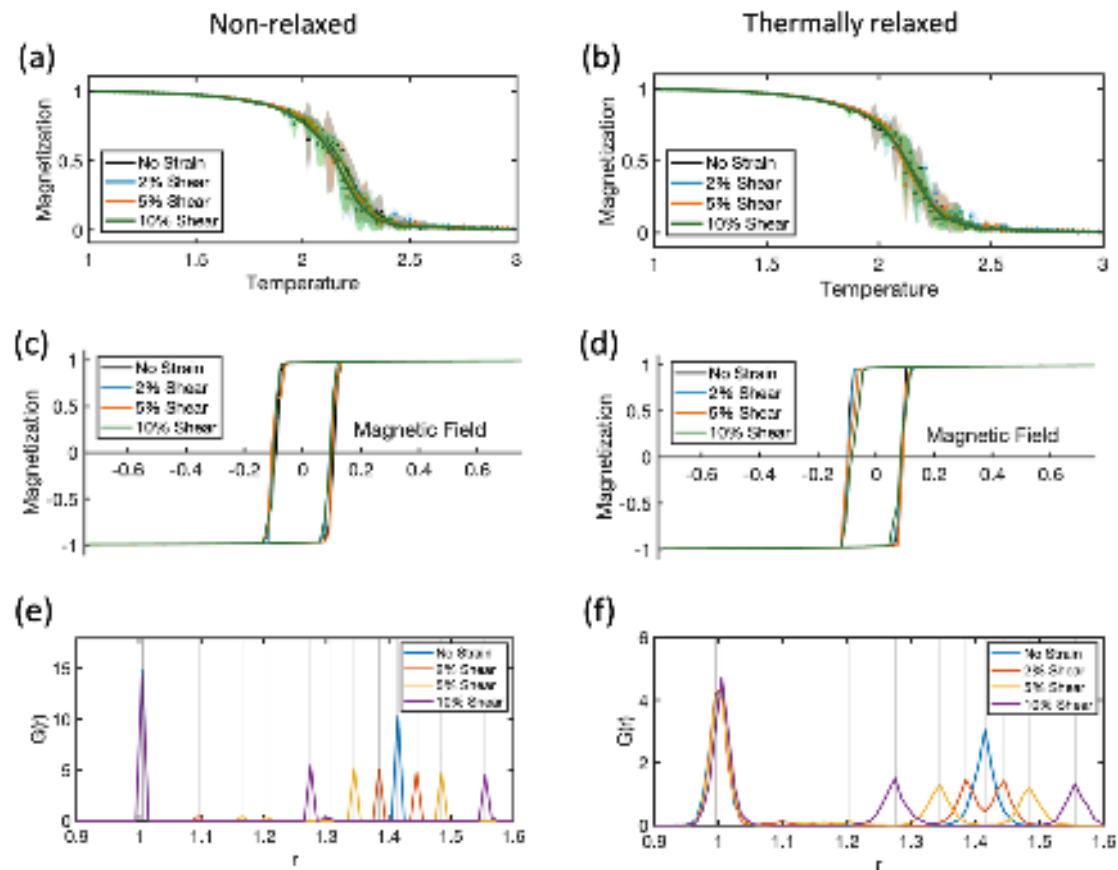

*Figure 3* Analysis of magnetic properties under pure shear deformation. (a) and (b) show the magnetization versus temperature curves for the non-relaxed and thermally relaxed systems, respectively (the shaded areas represent the standard deviation over 10 simulations). (c) and (d) show the magnetization versus magnetic field hysteresis curves, demonstrating typical ferromagnetic hysteresis across all pure shear strains. (e) and (f) Radial distribution function G(r) at T=0.1. (each plot (a-d) represents the average results from 10 simulations.).

Figure 3 illustrates the influence of various pure shear strains (2, 5, and 10%) on magnetization. The magnetization versus temperature graphs for both the non-relaxed (a) and thermally relaxed (b) conditions showed that the transition temperature was minimally influenced by the shear strain. The plots of magnetization vs. magnetic field (c and d) exhibit hysteresis behavior, with negligible variations observed among the different shear strains. This indicates that the magnetization process was not considerably affected by the shear strain under these conditions. The radial distribution function, commonly denoted as G(r), is an important concept for analyzing the arrangement of atoms in crystals, amorphous solids, and liquids. It offers data on the likelihood of encountering a pair of particles that are a certain distance from the predicted likelihood for a perfectly random distribution with the same density. In 2D, the radial distribution function G(r) is defined as

$$G(r) = \frac{1}{\rho} \frac{dN(r)}{2\pi r \, dr} \qquad (4)$$

where ρ is the average particle density and dN(r) is the number of particles located within an annulus of radius r and thickness dr centered around a reference particle. The G(r) plots in Figure 3 (e) and (f) show that pure shear deformation has a lesser impact on the distance between adjacent neighbors than the distance between the next-nearest neighbors; therefore, it does not generate a disorder.

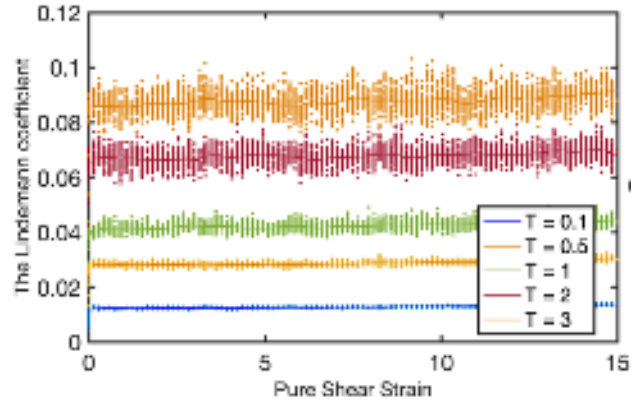

**Figure 4** Lindemann coefficient as a function of shear strain at five different temperatures: T = 0.1, T = 0.5, T = 1, T = 2, and T = 3 (each dot is a simulation result at a specific strain, and the solid line is the mean fit of the data.).

The Lindemann coefficient is an important metric for investigating disorders. The Lindemann criterion states that a crystal undergoes a transition to disorder when the root mean square (RMS) displacement of atoms from their equilibrium positions reaches a specific fraction of the interatomic distance. This criterion provides a numerical assessment of the level of disorder in a material with respect to the distance between atoms[27]. The Lindemann coefficient can be expressed as

$$C_L = \frac{\langle u^2 \rangle^{1/2}}{a} \tag{5}$$

where $\langle u^2 \rangle^{1/2}$ is the root mean square (RMS) atomic displacement and a is the lattice constant. The Lindemann coefficient exhibited a horizontal stable value with a temperature-dependent width throughout the range of shear strains (Figure 4). This indicates that the material retained its structural integrity and stability, even when subjected to higher levels of shear deformation. Consequently, when the temperature was low, the atomic displacements were negligible and the material maintained its organized arrangement. Figure 4 illustrates the correlation between the temperature and structural stability of the 2D materials subjected to pure shear strain. This shows that higher temperatures intensify atomic displacements and lead to increased structural instability. However, the average value remained consistent across the entire strain range and exerted minimal influence on the magnetic properties.

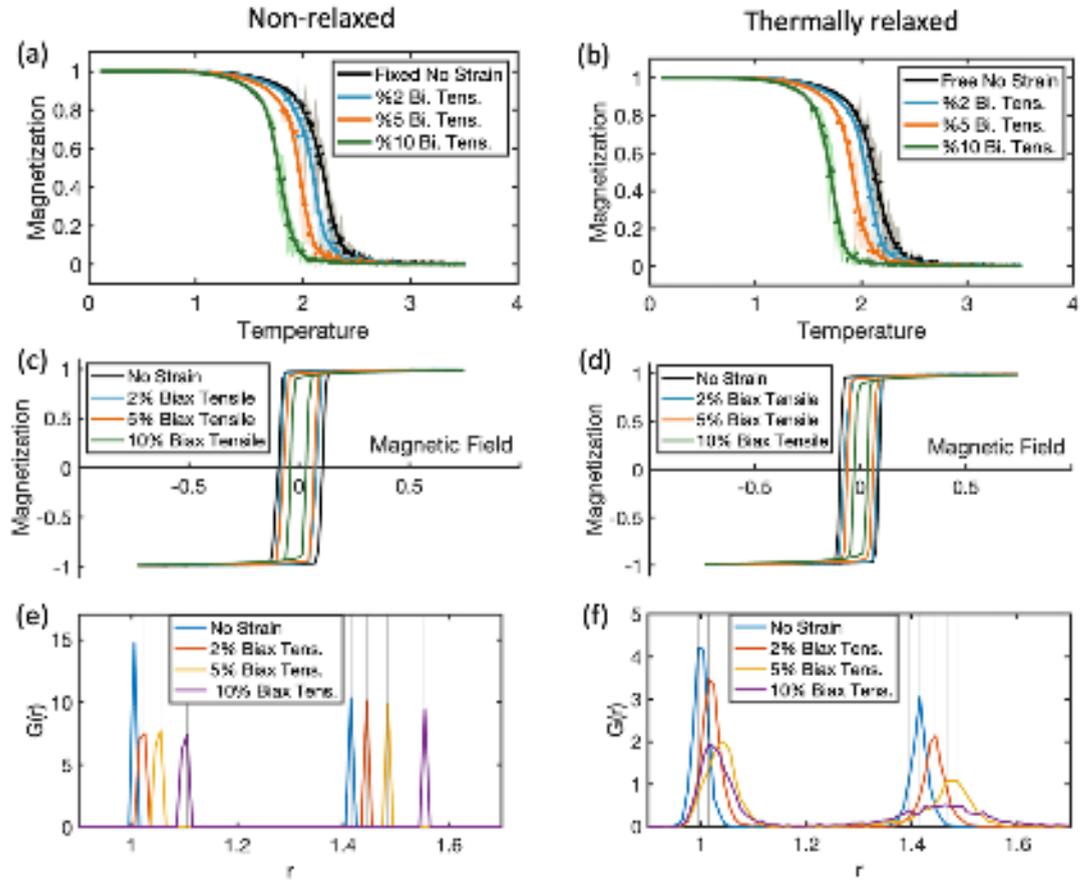

**Figure 5** *Analysis of magnetic properties under tensile deformation in 2D materials. (a) and (b) show the magnetization versus temperature curves for the non-relaxed and thermally relaxed systems, respectively (the shaded areas represent the standard deviation over the 10 simulations). . (c) and (d) show the magnetization versus magnetic field hysteresis curves, demonstrating typical ferromagnetic hysteresis across all tensile strains. (e) and (f) Radial distribution function G(r) at T=0.1. ( Each plot (a-d) represents the average results from 10 simulations.).*

The influence of biaxial tensile strain on magnetization was thoroughly examined in both non-relaxed and thermally relaxed states (Figure 5). The transformation temperature decreases when the biaxial tensile strain increases. In addition, thermally relaxed systems exhibited lower transformation temperatures. Additionally, the hysteresis loops of magnetization versus magnetic field indicate that higher tensile strains lead to a decrease in coercivity, with thermally relaxed systems showing an approximately 33% reduction (from 0.045 to 0.03) in coercivity compared with non-relaxed systems. The radial distribution function analysis at T=0.1 shows that thermal relaxation causes broader G(r) peaks with lower top values, and a notable shift of the tensile peak back towards the ideal no-strain position, attributed to cryogenic tearing.

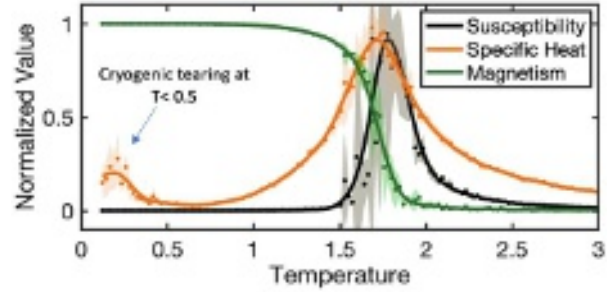

*Figure 6* Temperature dependence of normalized susceptibility, specific heat, and magnetism (shaded areas represent the standard deviation across 10 simulations, dots indicate the mean values at specific temperatures, and solid lines correspond to fitted functions).

Figure 6 illustrates the normalized values of susceptibility, specific heat, and magnetism as functions of temperature, providing a detailed view of the phase transition of the material. The susceptibility curve shows a prominent peak just above T = 1.8, indicating a critical point at which the magnetic response of the material is significantly enhanced and a phase transition occurs. Specific heat displays two notable peaks, one around T = 1.75, associated with a major phase transition, and a smaller one at T < 0.5, labeled as "Cryogenic tearing," indicating an anomaly at very low temperatures. These observations are characteristic of systems with coupled magnetic and elastic properties, such as those modeled by the Ising model with magnetoelastic interactions, where temperature-induced changes in structural and magnetic ordering lead to complex behavior [28].

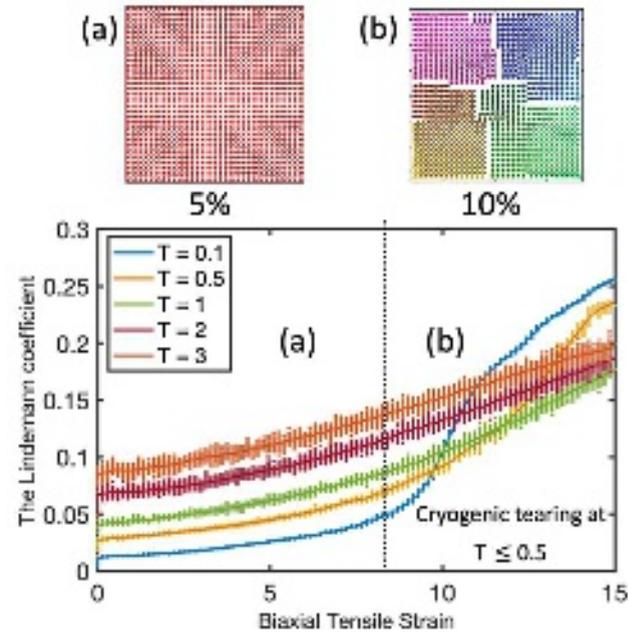

*Figure 7* The relationship between Lindemann Coefficient and biaxial tension strain (Each point is a simulation result at specific strain and solid line is the mean fit of the data.). (a) 5% tensile strain, (b) 10% tensile strain (domains are colored by motion direction and bond length)

Figure 7 illustrates the relationship between the Lindemann coefficient and the biaxial tension strain in a strained 2D magnetic material system at different temperatures. The top panels (a) and (b) show the magnetic system at 5% and 10% biaxial tensile strains, respectively. Panel (a) shows a uniform alignment of magnetic vectors, indicative of an ordered state, and panel (b) displays domain formation, suggesting a transition to a disordered state. The graph reveals that, as the tensile strain increases, the Lindemann coefficient also increases, reflecting enhanced atomic fluctuations owing to both strain and thermal energy. Notably, at lower temperatures (T ≤ 0.5), the Lindemann coefficient sharply increases after 8% strain, reaching the 0.2-0.25 range, indicative of a disorder caused by cryogenic tearing, as illustrated in panel (b). At higher temperatures (T > 0.5), the coefficient converged to the 0.17-0.2 range. This behavior underscores the critical role of strain in influencing the structural integrity and order of the system.

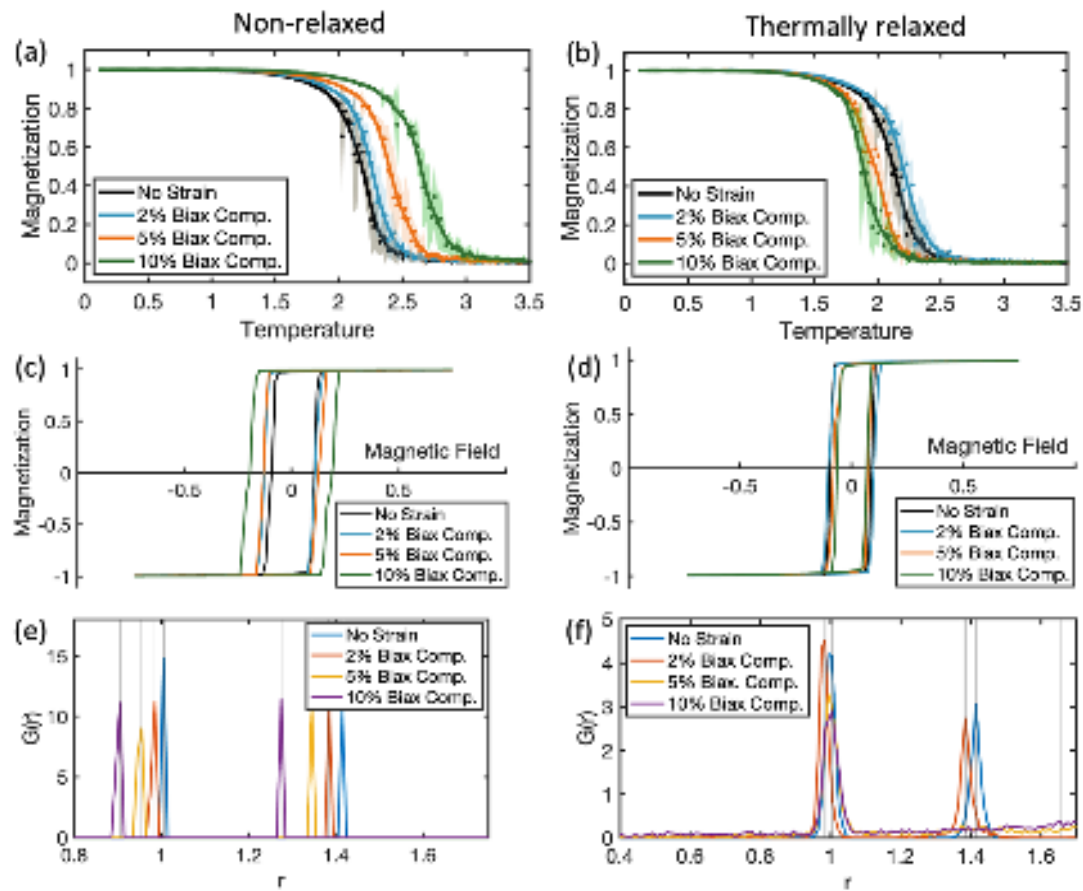

*Figure 8* Analysis of magnetic properties under compressive deformation. (a) and (b) show the magnetization versus temperature curves for the non-relaxed and thermally relaxed systems, respectively (the shaded areas represent the standard deviation over 10 simulations). (c) and (d) show the magnetization versus magnetic field hysteresis curves for non-relaxed and thermally relaxed systems, respectively, demonstrating typical ferromagnetic hysteresis across all compressive strains. (e) and (f) Radial distribution function G(r) at T=0.1. ( Each plot (a-d) represents the average results from 10 simulations.).

Figure 8 illustrates the impact of biaxial compression on the magnetization of a system under various conditions, including non-relaxed and thermally relaxed states. Magnetization as a function of temperature is displayed in the top row for different amounts of biaxial compression (0, 2, 5, and 10%). Under non-relaxation conditions, the application of more biaxial compression results in a wider and shifted transition temperature. This observation suggests that strain has a notable impact on the phase transition. A system in a state of thermal relaxation displays intricate patterns with slightly more distinct transitions, indicating that thermal relaxation aids in reducing the widths of transitions. At low strain values (less than 3%), the transformation temperature increased, similar to the ideal condition. However, at high strain levels, the transformation temperature decreased significantly. In addition, under a high strain, the magnetic behavior of thermally relaxed 2D materials converges to a narrow region. The middle row shows magnetization as a function of the magnetic field at a fixed temperature. For the non-relaxed cases, higher biaxial compression levels resulted in broader hysteresis loops, implying that strain enhanced the coercive field and overall magnetic stability. At a compressive strain of 5-10%, the order of the next-nearest neighbors was entirely disrupted (Figure 8(f)).

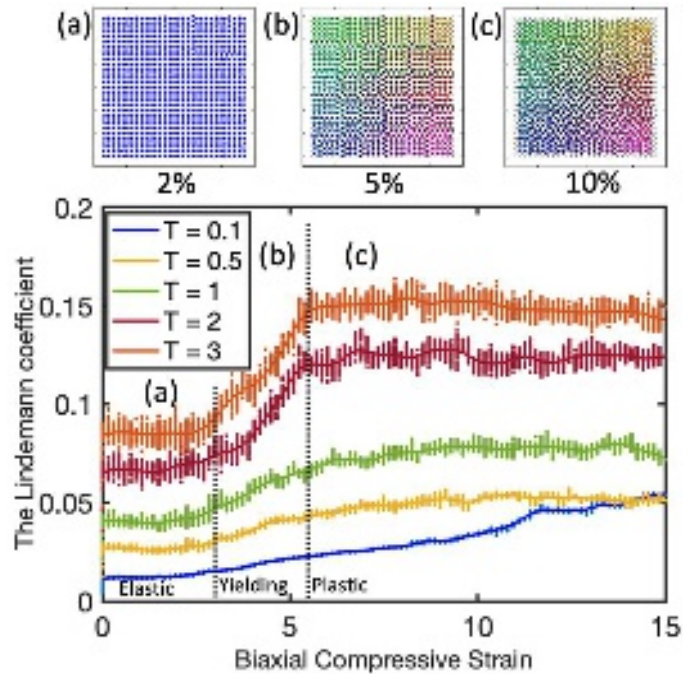

*Figure 9* Lindemann Coefficient vs. biaxial compressive strain (each dot is a simulation result at a specific strain, and the solid line is the mean fit of the data.). (a) 2%, (b) 5%, and (c) 10% biax Compressive strain. (domains are colored by motion direction and bond length)

Figure 9 illustrates the correlation between the Lindemann coefficient and the biaxial compressive strain at different temperatures (T = 0.1, 0.5, 1, 2, and 3). This reveals the distinct stages of the material response to the compressive strain. The Lindemann coefficient remained consistent across a wide range of temperatures, exhibiting a minimal variation in response to a strain of up to 3%. This suggests that even a small amount of strain can significantly affect material structure. The Lindemann coefficient exhibited a distinct increase above a 3% strain, reaching a maximum at strains exceeding 6%. An increase in the degree of entropy indicated a transition from the elastic region to the yielding region. The Lindemann coefficient reached its maximum value when the strain reached 6%. Subsequently, it remained unchanged, indicating that the material had entered the plastic zone and experienced substantial structural disorder. This loss of long-range order, indicative of amorphization, implies a transition from a crystalline to an amorphous state, as evidenced by the disappearance of the next-nearest neighbor peaks in the G(r) plots (Figure 8 (f)). At T = 0.1, the Lindemann coefficient deviates from the overall pattern and experiences a slight increase of 12% in strain followed by heightened volatility. This suggests that there was greater variation in the movement of atoms. This disparity indicates that the deformation reaction of the material becomes highly complex at extremely low temperatures. Insets (a), (b), and (c) provide visual evidence of the following phases: when subjected to a strain of 2%, (a) exhibited minor distortion. Domain formation was observed when the strain increased to 5% (b). At a strain of 10%, (c) displays a significantly disordered structure. These findings indicate that the material maintains its organization during low-strain compression. However, once the strain reached approximately 3%, there was a noticeable increase in the level of disorder, reaching its maximum value at a strain of approximately 6%.

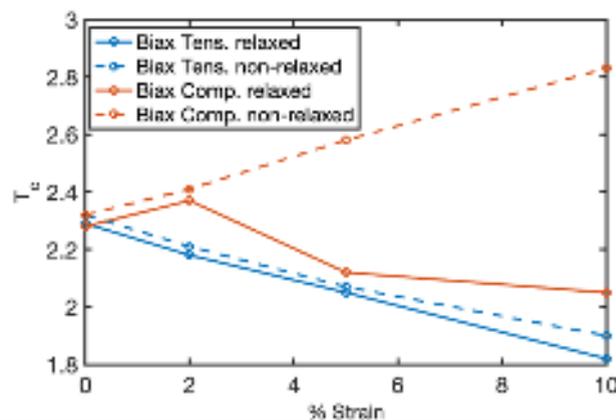

**Figure 10** *Variation in critical temperature ($T_c$) as a function of biaxial tensile and compressive strain (% strain) for relaxed and non-relaxed systems.*

Figure 10 illustrates $T_c$ as a function of strain percentage under biaxial tensile and compressive conditions. The data reveal that the biaxial tensile strain leads to a decrease in $T_c$. Conversely, the biaxial compressive strain exhibited an initial increase in $T_c$ under thermal conditions followed by a subsequent decrease. However, it remains higher than the overall tensile strain. These findings align with those of previous studies by Gao et al. (2016), who demonstrated structural changes in GaN sheets under biaxial compression[29], and Juntree et al. (2023), who observed changes in the magnetic properties of MnBi under compression [30]. The most significant finding from the plot was the drastic deviation in $T_c$ under biaxial compression after the elastic region.

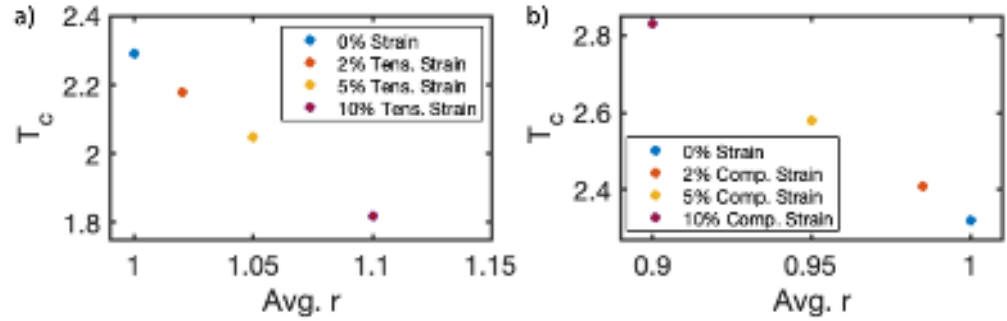

*Figure 11 Critical temperature versus interatomic distance (avg. r) for non-relaxed systems under tensile (a) and compressive (b) strains.*

Figure 11 shows the relationship between the critical temperature ($T_c$) and the average radius (avg. r) for non-relaxed systems, exhibiting a near-linear response under both tensile and compressive strains. This trend is attributed to variations in bond length, which is the sole parameter affecting the system. Specifically, under compressive strain, a decrease in the bond length results in an enhancement in the magnetic coupling strength, thereby increasing $T_c$. Conversely, under tensile strain, an increase in the bond length leads to a reduction in the magnetic coupling strength, consequently decreasing $T_c$ as the strain increases.

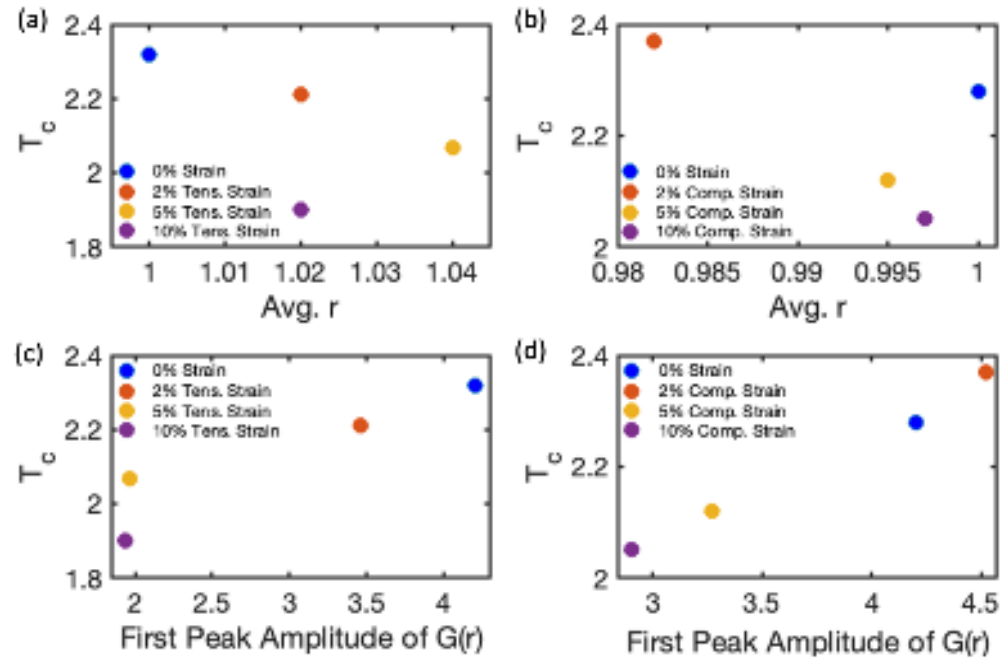

*Figure 12 Critical temperature ($T_c$) as a function of average interatomic distance (avg. r) and the first peak amplitude of G(r) under different strain conditions for the relaxed systems. (a) and (b) show the relationship between $T_c$ and the average interatomic distance r for tensile (2%, 5%, 10%) and compressive (2%, 5%, 10%) strains, respectively. (c) and (d) illustrate the correlation between $T_c$ and the first peak amplitude of G(r) for tensile and compressive strains, respectively. The data points reflect how varying the strain levels influence the atomic structure and the resulting magnetic phase transitions in the material.*

For relaxed systems, as the tensile strain increased to 5%, the average interatomic distance (avg. r) increased, leading to a significant reduction in $T_c$. This reduction results from the weakening of the exchange interactions as the atoms are pulled apart, thereby diminishing the magnetic coupling strength and consequently lowering the Tc. In addition, the first peak amplitude of G(r) decreased, indicating a loss of structural coherence as the material stretched, which further correlated with the observed decline in $T_c$. These mechanisms mutually reinforce each other. However, at 10% strain, cryogenic tearing occurred, where the average interatomic distance decreased but the first peak amplitude of G(r) remained constant. In this scenario, the reduction in Tc is attributed to the formation of separated domains due to tears (Figure 12(a),(c), and Figure 7(b)). In contrast, compressive strain decreased the average interatomic distance and increased the first peak amplitude of G(r) within the elastic region (<3% compressive strain), initially bringing atoms closer together and enhancing exchange interactions, leading to an increase in the critical temperature. However, as compression intensifies, the system undergoes structural instabilities such as buckling and twinning. Although these instabilities resulted in an increase in the avg. r, they also cause the first peak amplitude of G(r) to broaden and its amplitude to decrease, thereby disrupting the magnetic order and leading to a reduction in $T_c$ at higher strain levels (Figure 12(b) and (d)).

In summary, the effects of compressive and tensile strains on the atomic structure and magnetic properties of 2D materials highlight the complex interplay between strain-induced structural changes and magnetic behavior, whereas shear deformation appears to exert a minimal impact on their magnetic properties. A low compressive strain enhances the atomic interactions and structural order, as evidenced by the high first-peak amplitude of G(r) and a stable Lindemann coefficient. However, as the compressive strain increases beyond the elastic limit, the material undergoes structural shifts, such as twinning and buckling, leading to increased disorder and a decrease in the critical temperature, as reflected in the rising Lindemann coefficient. Conversely, tensile strain aligns the atomic structure more rigidly along the direction of the applied force, reducing configurational freedom but weakening atomic bonds, thereby increasing susceptibility to brittle failure. The resulting formation of domains owing to tearing and increased disorder contributed to a further decrease in the critical temperature at high tensile strains. This was indicated by an increase in the Lindemann coefficient, decrease in the average atomic distance, and reduction in the first peak amplitude of G(r). These findings highlight the importance of understanding strain effects in the design and optimization of 2D materials for advanced applications.

**Conclusion**

This study thoroughly examined the influence of mechanical strain on the magnetic characteristics of 2D materials using a dynamic Ising model. By investigating various strain scenarios, including biaxial tensile, compressive, and pure shear stresses, valuable insights into the atomic arrangement, magnetic property alterations, and resilience of 2D magnetic materials under different strain conditions were provided.

A key finding is that domain development can occur, even in this simple model. Specifically, cryogenic tearing under tensile strain does not affect magnetic behavior because of its occurrence at extremely low temperatures. In contrast, domains formed under compressive strain significantly influence magnetic behavior because they can be formed at any temperature. Pure shear deformation does not create domains or disorders, thereby minimally affecting magnetic behavior.

$T_c$ of 2D materials exhibits a distinct behavior under different strain conditions. The biaxial tensile strain generally reduces $T_c$, whereas the biaxial compressive strain initially increases $T_c$ but then decreases at higher strain levels. These findings are consistent with those of Gao et al. (2016), who demonstrated structural changes in GaN sheets under biaxial compression[29], and Juntree et al. (2023), who observed changes in the magnetic properties of MnBi under compression[30].

In summary, this study offers a comprehensive understanding of the complex relationship between the mechanical strain and the magnetic properties of 2D materials. These results are crucial for the development of strain-engineered 2D magnetic materials with characteristics tailored for advanced technological applications. Future research should focus on integrating the bond angle dependency and frustrated systems to further refine our understanding of the effects of strain on the magnetic properties.